%% file: 31st_December.tex
\documentclass[aps,twocolumn,nofootinbib,preprintnumbers,citeautoscript]{revtex4-1}
\usepackage{amsmath,amssymb,epsfig}
\usepackage{url}
\usepackage{enumerate}
\usepackage{mathpazo}
\usepackage{braket}
\usepackage{color}
\usepackage[colorlinks=true,citecolor=blue,urlcolor=blue]{hyperref}
\newcommand{\ketbra}[2]{|#1\rangle \langle #2|}

\newcommand{\tr}{\operatorname{Tr}}
\newcommand{\be}{\begin{equation}}
\newcommand{\ee}{\end{equation}}
\newcommand{\ba}{\begin{eqnarray}}
\newcommand{\ea}{\end{eqnarray}}

\def\be{\begin{equation}}
\def\ee{\end{equation}}

\def\be{\begin{equation}}
\def\ee{\end{equation}}

\def\beq{\begin{eqnarray}}\def\eeq{\end{eqnarray}}

\usepackage{graphicx}

\begin{document}

\title{Direct determination of entanglement monotones for arbitrary dimensional bipartite states using statistical correlators and one set of complementary measurements }

\author{Debadrita Ghosh$^{1,\dagger}$, Thomas Jennewein$^{2}$, Urbasi Sinha$^{1}$}
\email{usinha@rri.res.in}
\affiliation{$^{1}$Light and Matter Physics, Raman Research Institute, Bengaluru-560080, India}
\altaffiliation{now at Institut für Nanophotonik Göttingen e. V., Georg August Universit\" at G\" ottingen, Germany}
\affiliation{$^{2}$Institute for Quantum Computing, University of Waterloo, 200 University Avenue West, Waterloo, ON Canada}


\begin{abstract}

Higher dimensional quantum systems (qudits) present a potentially more efficient means, compared to qubits, for implementing various information theoretic tasks.  One of the ubiquitous resources in such explorations is entanglement. Entanglement Monotones (EMs) are of key importance, particularly for assessing the efficacy of a given entangled state as a resource for information theoretic tasks. Till date, investigations towards determination of EMs have focused on providing their tighter lower bounds. There is yet no general scheme available for direct determination of the EMs. Consequently, an empirical determination of any EM has not yet been achieved for entangled qudit states. The present paper fills this gap, both theoretically as well as experimentally. First, we derive analytical relations between statistical correlation measures i.e. Mutual Predictability ($\mathcal{MP}$), Mutual Information ($\mathcal{MI}$) and Pearson Correlation Coefficient ($\mathcal{PCC}$) and standard EMs i.e. Negativity ($\mathcal{N}$) and Entanglement of Formation ($\mathcal{EOF}$) in arbitrary dimensions. As a proof of concept, we then experimentally measure $\mathcal{MP}$, $\mathcal{MI}$ and $\mathcal{PCC}$ of two-qutrit pure states and determine their $\mathcal{N}$ and $\mathcal{EOF}$ using these derived relations. This is a useful addition to the experimenter's toolkit wherein by using a limited number of measurements (in this case 1 set of measurements), one can directly measure the EMs in a bipartite arbitrary dimensional system. We obtain the value of $\mathcal{N}$ for our bipartite qutrit to be 0.907 $\pm$ 0.013 and the $\mathcal{EOF}$ to be 1.323 $\pm$ 0.022. Since the present scheme enables determination of more than one entanglement monotone by the same limited number of measurements, we argue that it can serve as a unique experimental platform for quantitatively comparing and contrasting the operational implications of entanglement monotones as well as showing their non-monotonicity for a  given bipartire pure qudit state. 

\end{abstract}
\maketitle

\newpage

Entanglement \cite{Schrodinger} is one of the pivotal features of quantum mechanics that has deep-seated implications like nonlocality \cite{bell} and is a crucial resource for quantum communication and information processing tasks \cite{ben,deu,benn,JL03,ek,ABG+07,bruk,BCM+10,PAM+10,NPS14}. 
While the two-dimensional (qubit) case continues to be widely studied in the context of various applications of quantum entanglement, it has been gradually recognised that higher dimensional entangled states can provide significant advantages over standard two-qubit entangled states in a variety of applications, like, increasing the quantum communication channel capacity \cite{bennett,WDF+05}, enhancing the secret key rate and making the quantum key distribution protocols more robust in the presence of noise \cite{bech,CBK02,BCE+03,SV10} as well as enabling more robust tests of quantum nonlocality by reducing the critical detection efficiency required for this purpose \cite{ver}. At this stage, the following question merits attention. Given an entangled state, how does one assess its efficacy for a given application? To this end, it is important to know how much entangled the state is. So far, all the relevant investigations towards quantifying entanglement have focused on providing the tighter lower bounds on EMs, mainly concerning the entanglement of formation. No general scheme is yet available for determining the EMs for the arbitrary dimensional entangled states, and no empirical estimation of any EM has been realised for any higher dimensional entangled state. The present paper fills this gap, both theoretically as well as experimentally.

In this work, we formulate a procedure to determine the standard entanglement monotones like Entanglement of Formation ($\mathcal{EOF}$) and Negativity ($\mathcal{N}$) for the pure arbitrary dimensional bipartite states using only one or two sets of joint measurements. This general scheme is based on the analytical relationships that we derive between the EMs 
and the observable standard statistical correlators viz. Mutual Predictability ($\mathcal{MP}$), Mutual Information ($\mathcal{MI}$) and Pearson Correlation Coefficient ($\mathcal{PCC}$). This procedure is then illustrated by a proof-of-concept experiment. We generate a highly pure bipartite photonic spatial-bin qutrit state by using the pump beam modulation technique that has been earlier devised by us \cite{Ghosh:18}, which is easily scalable to higher dimensions. We certify the entanglement by performing appropriate measurements in complementary basis as has been argued to be necessary for such certification \cite{PhysRevA.57.3123}. We then apply our derived relations between the $\mathcal{N}$ and $\mathcal{EOF}$ with $\mathcal{MP}$, $\mathcal{MI}$ and $\mathcal{PCC}$ for the prepared two-qutrit state. Hence, this constitutes the first direct experimental determination of the standard entanglement monotones, and that too, using only one set of joint local complementary measurements in the performed experiment, for the higher dimensional states. Furthermore, in a separate work \cite{theorypaper}, it has been shown how the scheme formulated here can be extended for different types of empirically relevant mixed bipartite entangled states.  

Next, we consider the implications of such a scheme for quantitative comparisons of entangled monotones. For example, one may consider the question whether these entanglement monotones are monotonic with respect to each other for the higher dimensional pure states. Here we may recall that in the pure bipartite qubit scenario, they are monotonic with respect to each other and choosing one entanglement monotone to compare the entanglement present in different states should be sufficient even if the different monotones do not capture the same amount of deviation of a given state from the maximally entangled state, as was shown in \cite{2019arXiv190709268S}.

In our experiment, since both the EMs given by $\mathcal{EOF}$ and $\mathcal{N}$ are determined independently for the prepared two-qutrit pure state, it provides a unique empirical means for not only quantitatively highlighting the difference in the operational meanings of these EMs in any specific context, but also for investigating their monotonicity. Importantly, this can be done using only one set of local joint complementary measurements. 

Further, our theoretical analysis in this context clearly brings out the hitherto unexplored feature of non-monotonicity between higher dimensional entanglement monotones, here specifically $\mathcal{EOF}$ and $\mathcal{N}$. Thus, for the higher dimensional entangled state, this line of study opens up the possibility of future experiments demonstrating such non monotonicity as well as its attendant implications.

Now, proceeding to the specifics of this paper, we begin with a background section where we briefly discuss the current status of entanglement certification and quantification in higher dimensional systems and introduce the three statistical correlation measures that we go on to use in our manuscript. Next, we derive and establish the relations between these statistical measures and two known entanglement monotones viz. $\mathcal{N}$ and $\mathcal{EOF}$, followed by description of the experimental scheme that we have used and the implications of the results obtained.

\section{Background}


The enterprise of experimentally realising higher-dimensional entangled states, along with studies of optimal certification and quantification of the entanglement has gained tremendous importance in recent times. 
While many approaches for characterization of higher dimensional entanglement \cite{SSV13,SSV14,SHR17,SSC17,TDC+17,MGT+17,BVK+17,SH18,Roy05,DAC17,HMK+16,MBM15} have been studied (see \cite{PRAus} for an overview), only few schemes provide both necessary and sufficient certification together with quantification of high-dimensional entanglement. The quantification schemes that have been suggested so far seem to have focused essentially on providing bounds on entanglement monotones.
The usual method of characterising quantum states, i.e., Quantum State Tomography (QST) and attendant estimation of any entanglement monotone from the QST data would require determination of an increasingly large number of independent parameters as the dimension of the system grows \cite{IOA17}. While this challenge has been addressed to certify entanglement by using an Entanglement Witness (EW), \cite{barbara} imperfections in a typical EW measurement may result in an inaccurate judgement about the entanglement present. To address such measurement imperfections, measurement-device-independent schemes have been developed. In a recent work, certification of entanglement has been shown with an EW protocol by quantifying a lower bound on a non-standard entanglement measure \cite{YuGuo}.\\
A different approach to quantify entanglement can be done by determining the Schmidt number \cite{SSV14,JSperling}. But again, the necessary number of measurements to determine the Schmidt number increases with the system-dimension. \\
Formulating  experimentally efficient methods for the characterization of higher-dimensional entangled states based on a limited number of measurements has become an active area of research \cite{TCA+14, gio,how,EKH17,natcom}. There are recent studies to certify entanglement by determining a lower bound to $\mathcal{EOF}$ from a very few local measurements \cite{BVK+17}. Notably, there are other interesting approaches to characterize entanglement using two mutually unbiased bases and the standard statistical correlators. Studies have shown that measuring such correlators w.r.t the mutually unbiased bases having an upper bound can reveal entanglement in a quantum state \cite{SHB+12,MBM15}. In the context of higher dimensions, these studies were restricted to providing a separability bound \cite{SHB+12} and providing a conjecture regarding such correlators certifying entanglement \cite{MBM15}. We take these ideas forward to a firmer theoretical footing by deriving analytic relations between such correlators and entanglement monotones in arbitrary dimensional pure states, thus providing a first empirically testable relation between the amount of entanglement and statistical correlation measures. We then go on to verify these relations experimentally for the three dimensional case, thus quantifying the amount of entanglement by performing a first direct measure of entanglement monotones.\\
Following the scheme given in \cite{SHB+12}, considering a $d\times d$ bipartite state, $\psi_{AB}$ with density matrix $\rho_{AB}$,
we define the three measures of statistical correlations of interest in terms of joint local measurement on the state.   $\mathcal{MP}$, $\mathcal{MI}$ w.r.t two measurement bases $\{\ket{a_i}\}$ and $\{\ket{b_i}\}$ for sub-systems A and B respectively and $\mathcal{PCC}$ for any joint observable $\hat{A} \otimes \hat{B}$ are defined as,
\begin{equation}
\label{bg1}
 \mathcal{MP}_{AB}=\sum\limits_{i=0}^{d-1}\braket{a_i, b_i | \rho_{AB} | a_i, b_i}   
\end{equation}
\begin{widetext}
\begin{align}
 \label{bg2}
 \mathcal{MI}_{AB} = \sum\limits_{i,j=0}^{d-1} \braket{a_i, b_j | \psi_{AB} | a_i, b_j} \log_2 \left(\frac{\braket{a_i, b_j | \psi_{AB} | a_i, b_j}}{\braket{a_i | \mathrm{Tr}_A\{\psi_{AB}\} | a_i}\braket{ b_j |\mathrm{Tr}_B\{\psi_{AB}\} | b_j}}\right)   
\end{align}
\end{widetext}
\begin{equation}
\label{bg3}
\mathcal{C}_{AB} = \frac{\langle \hat{A}\otimes \hat{B} \rangle-\langle \hat{A}\otimes \mathbb{I} \rangle\langle \mathbb{I}\otimes \hat{B} \rangle}{\sqrt{\langle \hat{A}^2 \otimes\mathbb{I}\rangle-\langle \hat{A}\otimes\mathbb{I}\rangle^{2}}\sqrt{\langle \mathbb{I}\otimes \hat{B}^2\rangle-\langle\mathbb{I}\otimes \hat{B}\rangle^{2}}} 
\end{equation} respectively. 


\section{Relating Mutual Predictability with Negativity for arbitrary dimensional pure bipartite state}
We now derive the relation between $\mathcal{MP}$ and $\mathcal{N}$ in arbitrary dimension $d$. 
Consider a pure bipartite state written in the Schmidt decomposition
\begin{equation}
 \label{mp1}   
 \psi_{AB}=  \sum_{i=0}^{d-1} \lambda_i \ket{i_A,i_B}
\end{equation} with density matrix,
\begin{equation}
\label{density}
\rho_{AB}=\sum\limits_{i,j=0}^{d-1} \lambda_i \lambda_j \ket{i_A,i_B}\bra{j_A,j_B}
\end{equation}
where $\{\ket{i_A,i_B}\}$ is the Schmidt basis.
When measurement is done w.r.t the $k_{th}$ mutual unbiased basis (MUB), $\mathcal{MP}$ as defined in Eq. (\ref{bg1}) is written as,
\begin{equation}
\label{mp2}
\mathcal{MP}_k=\sum_{m=0}^{d-1}\, \sum_{i,j=0}^{d-1} \lambda_i \lambda_j  \braket{\psi_m^k ,\phi_m^k \vert i_A, i_B} \braket{j_A,j_B \vert \psi_m^k,\phi_m^k}
\end{equation}
where $\{\ket{\psi_m}\}$ and  $\{\ket{\phi_m}\}$ are the measurement bases for subsystems A and B respectively and the corresponding MUB is $\ket{\psi_m^k,\phi_m^k}$. The maximum value of $\mathcal{MP}$ for measurements done w.r.t the $k_{th}$ MUB is given in Eq.(\ref{mp3}). Details of this derivation may be found in Appendix \ref{sectionA}.
\begin{equation}
\label{mp3}    
\mathcal{MP}_k=\sum\limits_{i,j=0}^{d-1} \frac{\lambda_i \lambda_j}{d}=\frac{1+2\mathcal{N}}{d}
\end{equation} where $\mathcal{N}$ for pure bipartite state defined as, $\mathcal{N}=\frac{1}{2} \sum\limits_{i,j=0;i\neq j}^{d-1} \lambda_i \lambda_j$ \cite{ETS15}.

$\mathcal{MP}$ w.r.t the measurement basis is maximum when $\ket{\psi_i}=\ket{i_A}$ and $\ket{\phi_j}=\ket{j_B}$ i.e. for the Schmidt bases and is given by Eq.(\ref{mp4}).
\begin{equation}
\label{mp4}
\mathcal{MP}=\sum_{m=0}^{d-1}\sum_{i,j=0}^{d-1}\lambda_i\lambda_j\delta_{mi}\delta_{mj}=\sum_{i,j=0}^{d-1}\lambda_i\lambda_j\delta_{ij}=1
\end{equation}
Combining Eqs. \ref{mp3} and \ref{mp4} the sum of $\mathcal{MP}$ is maximized and turns out to be,
\begin{equation}
\label{mp5}
\mathcal{MP}_{max}=1+\frac{1+2\mathcal{N}}{d}
\end{equation}
Details of this derivation may be found in Appendix \ref{sectionA}. Here one may just briefly remark that for deriving the value of $\mathcal{N}$ from the measured value of $\mathcal{MP}$, it is empirically advantageous to consider single $\mathcal{MP}$ because then the associated error range is less than when the sum of $\mathcal{MP}$s is considered. \\
To verify the efficacy of the relation given by Eq. (\ref{mp3}) we perform a numerical calculation for the pure state, $\psi_{AB}$  with many random sets of $\{\lambda_i\}$ and plot its $\mathcal{N}$ against the numerically calculated $\mathcal{MP}$ and compare with Eq. (\ref{mp3}). For this we choose our measurement basis,
\begin{align}
    \label{eq:z_basis}
   & \{\ket{i,i}~ \forall~ i~\in~\{0,\cdots,d-1\}\} \\ 
    \label{eq:x_basis}
    &\left\{\sum\limits_{j=0}^{d-1} \omega_d^{-ij} \ket{j} \otimes \sum\limits_{j'=0}^{d-1} \omega_d^{ij'} \ket{j'^*}~ \forall~ i~\in~\{0,\cdots,d-1\}\right\}
\end{align}
Note that if $\ket{i}$ is the computational basis, the measurement bases are the eigenbases of the generalised $d-$dimensional Pauli $\sigma_z$ and $\sigma_x$ operators. The plots of $\mathcal{N}$ vs ${\mathcal{MP}}$ are shown in Fig. 1.
\begin{figure}
\centering
\includegraphics[scale=0.6]{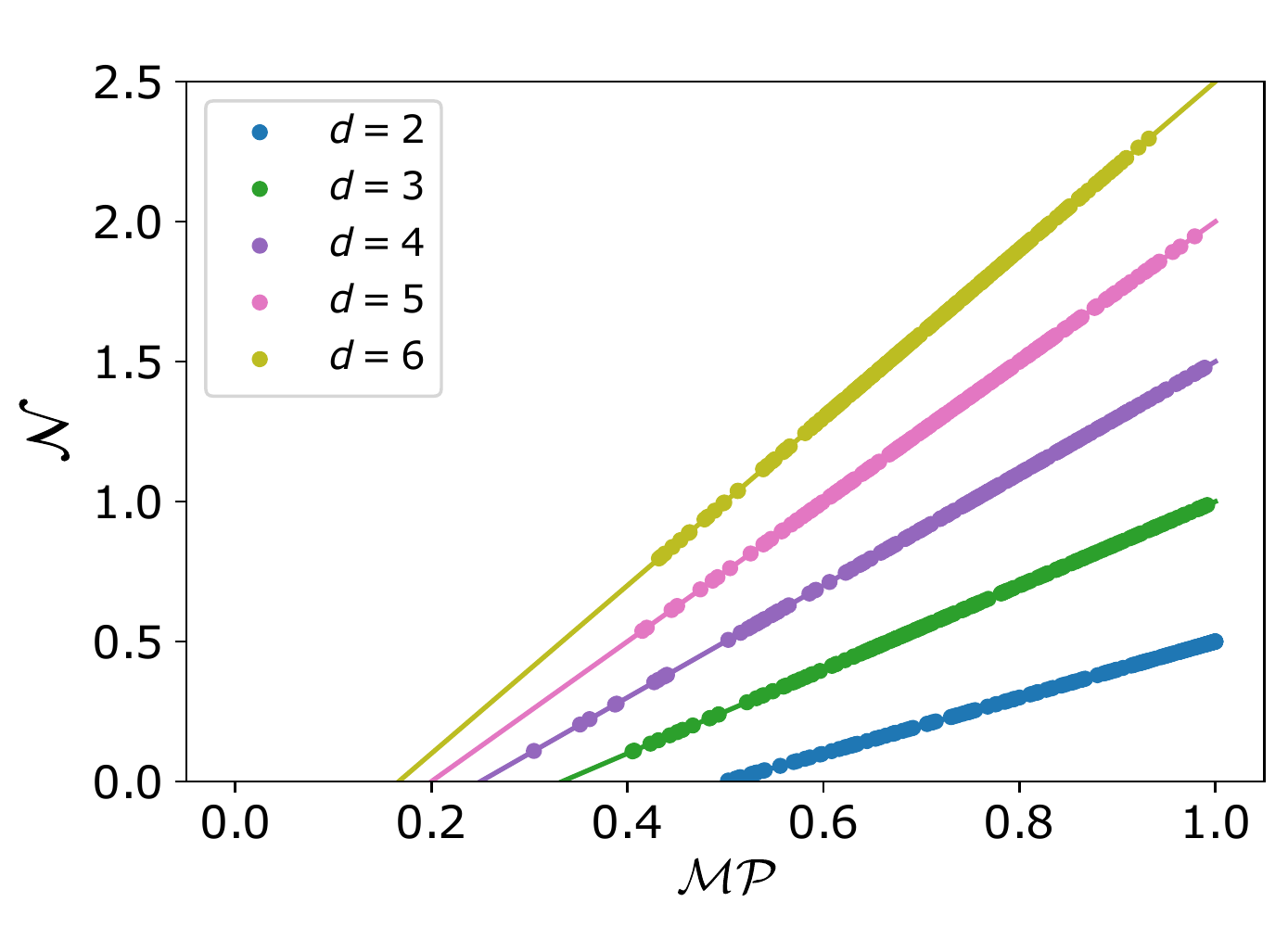}
\caption{Negativity vs Mutual Predictability w.r.t.~basis as shown in Eq. (\ref{eq:x_basis}), for a bipartite pure state for different dimensions. The solid lines indicate the empirical relation Eq. (\ref{mp4}) and the points are generated from numerical calculation.}
\label{}
\end{figure}\\

\section{Deriving an analytic relation between $\mathcal{PCC}$ and  $\mathcal{N}$ for pure bipartite state}

It is noteworthy that $\mathcal{PCC}$ has  so far been used in physics only in limited contexts \cite{BGP+17, BKO02} until recently when Maccone et al. \cite{MBM15} suggested its use for entanglement characterization. Let us suppose two spatially separated parties, Alice and Bob, share a bipartite pure or mixed state in an arbitrary dimension; Alice performs two dichotomic measurements $A_1$ and $A_2$ and Bob performs two dichotomic measurements $B_1$ and $B_2$ on their respective subsystems. Then, Maccone et al. conjectured that the sum of two $\mathcal{PCC}$s being greater than $1$ for appropriately chosen mutually unbiased bases would certify entanglement of bipartite systems, i.e., for $A_1=B_1=\sum_j a_j \ketbra{a_j}{a_j}$ and $A_2=B_2=\sum_j b_j \ketbra{b_j}{b_j}$,
 \begin{equation}
 \label{pcc1}
|\mathcal{C}_{A_1B_1}|+|\mathcal{C}_{A_2B_2}|>1,
 \end{equation}
would imply entanglement. Here, $\{\ket{a_j}\}$ is mutually unbiased to $\{\ket{b_j}\}$.  However, Maccone at al. justified this conjecture only by showing its applicability for bipartite qubits  and the validity of this conjecture has remained uninvestigated for dimensions $d>2$. In this work, we justify the validity of this conjecture for pure bipartite qutrits by deriving an analytic relation between $\mathcal{PCC}$ and $\mathcal{N}$ for any arbitary dimension $d$ (details of this derivation are provided in a separate paper \cite{theorypaper}), and this relation is tested by applying it to quantify the amount of entanglement in our experimentally generated pure bipartite qutrits ($d=3$). \\

$\mathcal{PCC}$ for a joint observable $X_A \otimes X_B$ is given by Eq. (\ref{bg3}) where the observable, $X$ can be written as $X=\sum\limits_{i,j:\braket{i\vert j}=0}\ket j \bra i$ and is such that it projects each computational basis $\ket i$ to the superposition of all the remaining basis vectors of the complete set. So, consequently,
\begin{equation}
\label{pcc2}
\langle \psi_{AB}| X_A\otimes\mathbb{I}|\psi_{AB}\rangle=\langle \psi_{AB}|\mathbb{I}\otimes X_B|\psi_{AB}\rangle=0
\end{equation}
for the state, $\psi_{AB}$ as given by the Eq. (\ref{mp1}). Furthermore, it can be shown that as discussed in \cite{theorypaper},
\begin{equation}
\label{pcc3}
  X^2=4\left((d-1)\mathbb{I}+\frac{(d-2)}{2} X\right)  
\end{equation} Using Eq. (\ref{pcc3}) we obtain,
\begin{equation}
\label{pcc4}
\langle X_A^2\otimes \mathbb{I}\rangle=\langle \mathbb{I}\otimes X_B^2\rangle= 4(d-1)
\end{equation} The joint expectation as appeared in the numerator of Eq. (\ref{bg2}) can be calculated as,
\begin{equation}
 \label{pcc5}   
\langle \psi_{AB}| X_A\otimes X_B|\psi_{AB}\rangle=4 \sum\limits_{i,j=0}^{d-1} \lambda_i \lambda_j=8\mathcal{N}
\end{equation}
Using Eqs. (\ref{pcc2}), (\ref{pcc4}), (\ref{pcc5}) we can show that
\begin{equation}
\label{pcc6}
C_{X_A X_B}= \frac{2\mathcal{N}}{(d-1)}
\end{equation}
Using $Z_A=\sum i \ket{i_A}\bra i$ and $Z_B=\sum j \ket{j_B}\bra{j}$ 
\begin{equation}
 \label{pcc7}   
 C_{Z_A Z_B}=1
\end{equation}
Thus by combining Eqs. \ref{pcc6} and \ref{pcc7} one gets,
\begin{equation}
 \label{pcc8} 
 C_{Z_A Z_B}+C_{X_A X_B}=1+\frac{2\mathcal{N}}{(d-1)}
\end{equation}

\section{Relating Mutual Information with Entanglement of Formation}

Let the common basis  of measurement for the  pair of observables $A$ and $B$ pertaining to Alice and Bob 
be the computational basis. For this choice of measurements, substituting the values of joint probabilities and the marginal probabilities for the pure two-qudit state as expressed in Eq.( \ref{mp1})
where $0 \le \lambda_i \le 1$ and $\sum_i\lambda^2_i=1$.
 In the expression for $\mathcal{MI}$ \cite{MBM15} as given by Eq. (\ref{bg2}) 
\begin{equation}
 I_{AB}=-\sum^{d-1}_{i=0}\lambda^2_i\log_2\lambda^2_i \label{MI}
\end{equation}

Now, note that, Entanglement of Formation for bipartite pure states $\ket{\psi}_{AB}$
is equal to the von Neumann entropy of either of the reduced 
density matrices,\cite{von} i.e.,
 $\mathcal{E}(\ket{\psi}_{AB})=S(\rho_A)=S(\rho_B)$, here 
 $S(\rho)=-\tr[\rho \log_2\rho]$. For the general pure two-qudit state as 
 given by Eq. (\ref{mp1}), $\mathcal{EOF}$ is 
given by the following expression:
\be \label{eofPtQd}
\mathcal{E}(\ket{\psi_d})=-\sum_i \lambda_i^2 \log_2 \lambda^2_i.
\ee
since $S(\rho_A)=S(\rho_B)=-\sum_i \lambda_i^2 \log_2 \lambda^2_i$.

From Eqs. (\ref{MI}) and (\ref{eofPtQd}), it then follows that $\mathcal{MI}$ pertaining to the computational basis on both
sides equals the $\mathcal{EOF}$ for any pure bipartite qudit state.




\section{Experimental details}
\subsection{Experimental scheme}

Common choices for higher dimensional photonic systems include those based on exploiting the Orbital Angular Momentum degree of freedom of a single photon \cite{PhysRevLett.116.073601,Palacios:11,Fickler13642,PhysRevA.76.042302,PhysRevA.69.023811}, spatial degree of freedom by placing apertures or spatial light modulators in the path of down-converted photons \cite{PhysRevA.78.012307,PhysRevA.79.043817,PhysRevA.80.062102,PhysRevLett.94.100501,PhysRevA.57.3123,PhysRevA.69.042305} as well as time-bin qudits \cite{PhysRevA.92.033802,PhysRevLett.93.180502,PhysRevA.66.062304}. 
Recently, our group has demonstrated a novel technique for spatial qutrit generation based on spatially modulating the pump beam in spontaneous parametric down-conversion (SPDC) by appropriately placed triple slit apertures \cite{Ghosh:18}. This leads to direct generation of highly correlated bipartite qutrits from the SPDC process, which we call spatial-bin qutrits. This technique has been shown \cite{Ghosh:18} to be more efficient and robust, also leading to a more easily scalable architecture than what is achieved by the conventional method \cite{PhysRevA.86.012321} of placing slits in the path of down-converted photons. It is also noteworthy that harnessing the spatial degree of freedom has the following unique feature. On the one hand, by only changing the plane of measurement, different measurement bases may be realised while on the other hand, different eigen bases can be realised by changing the detector positions along the plane. In our experimental context, for instance, the image plane and the focal plane of the lenses L2 and L3 as shown in Fig. 2 represent generalized $\sigma_z$ and $\sigma_x$ measurement basis respectively. We discuss more details of these measurements later in this section as well as in Appendix \ref{sectionD}.\\
\begin{figure*}
\centering
\includegraphics[width=1\textwidth]{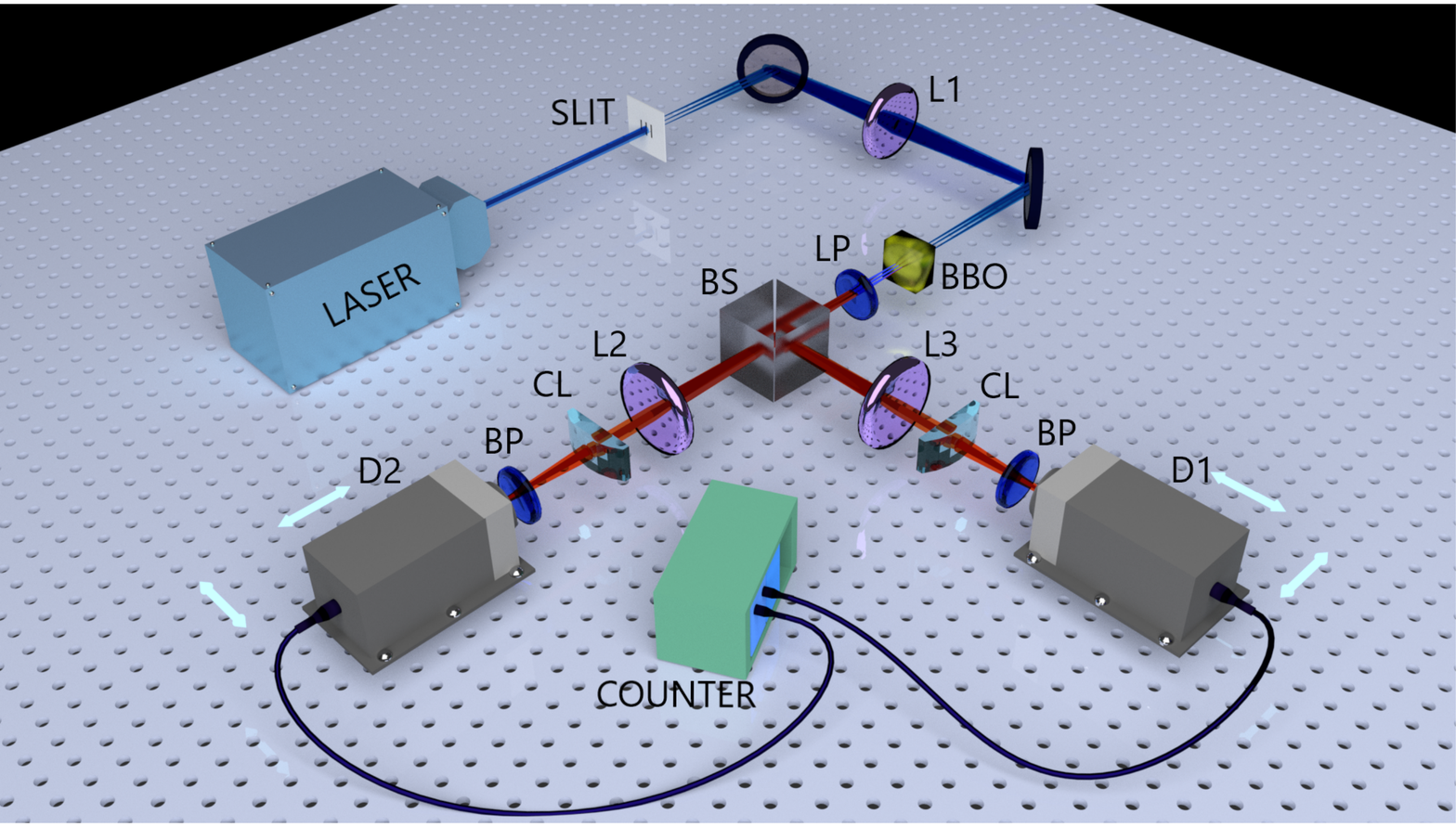}
\caption{Schematic of the experimental set-up. L1, L2, L3: Plano convex lenses, BBO: Nonlinear crystal for SPDC, LP: Long-pass filter, BS: 50-50 Beamsplitter, BP: Band-pass filter, CL: Cylindrical Lens, D1, D2: Single photon detectors. }
\label{}
\end{figure*}

\begin{figure}
\centering
\includegraphics[scale=0.300]{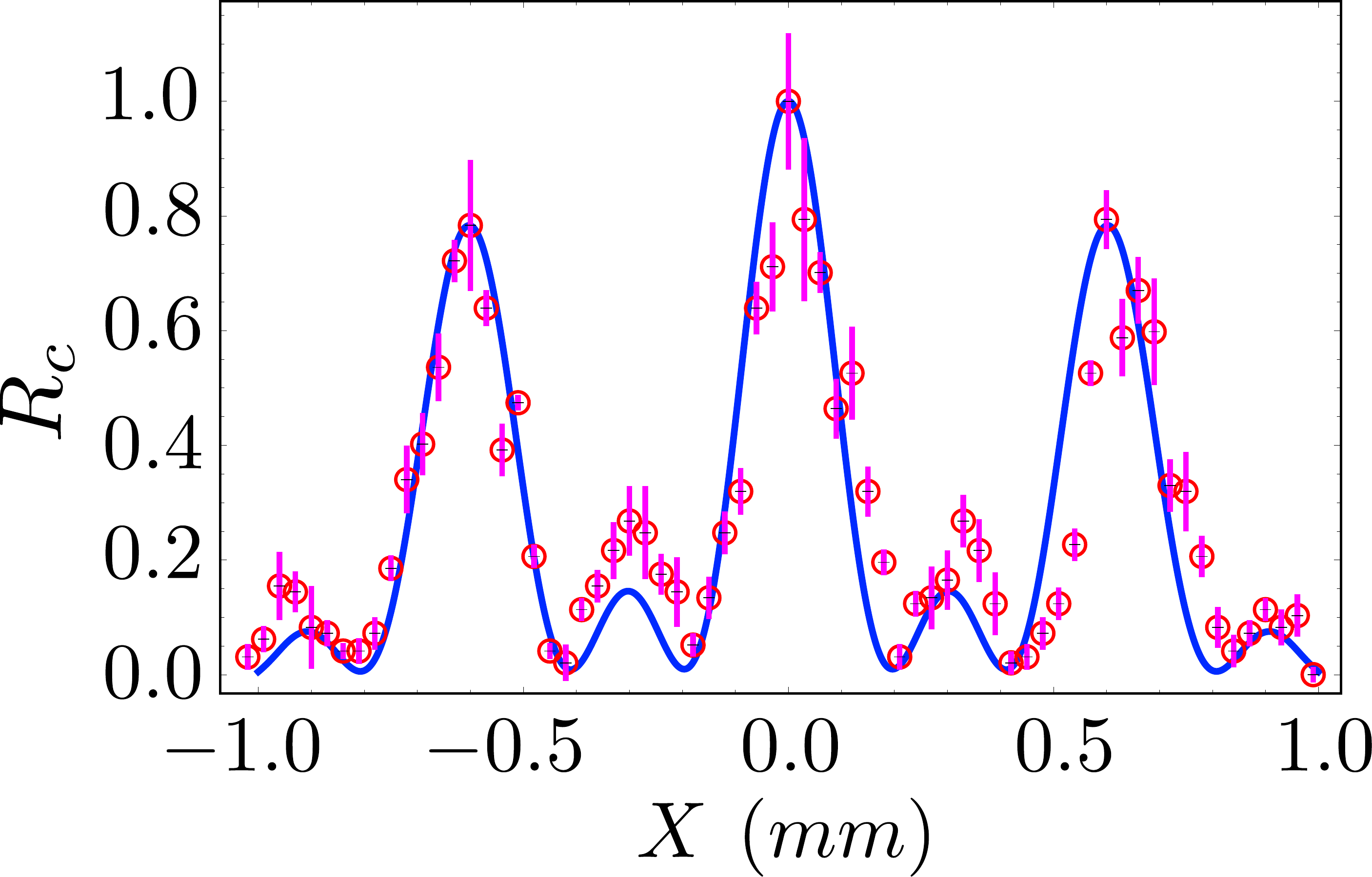}
\caption{Normalized coincidence count ($R_c$) vs detector position (X) in focal plane. Blue line indicates the theoretical prediction whereas the red circle indicates the experimental result. The normalised coincidence plot exhibiting interference is a certification of entanglement.}
\label{}
\end{figure}
\begin{figure*}
\includegraphics[width=1\textwidth]{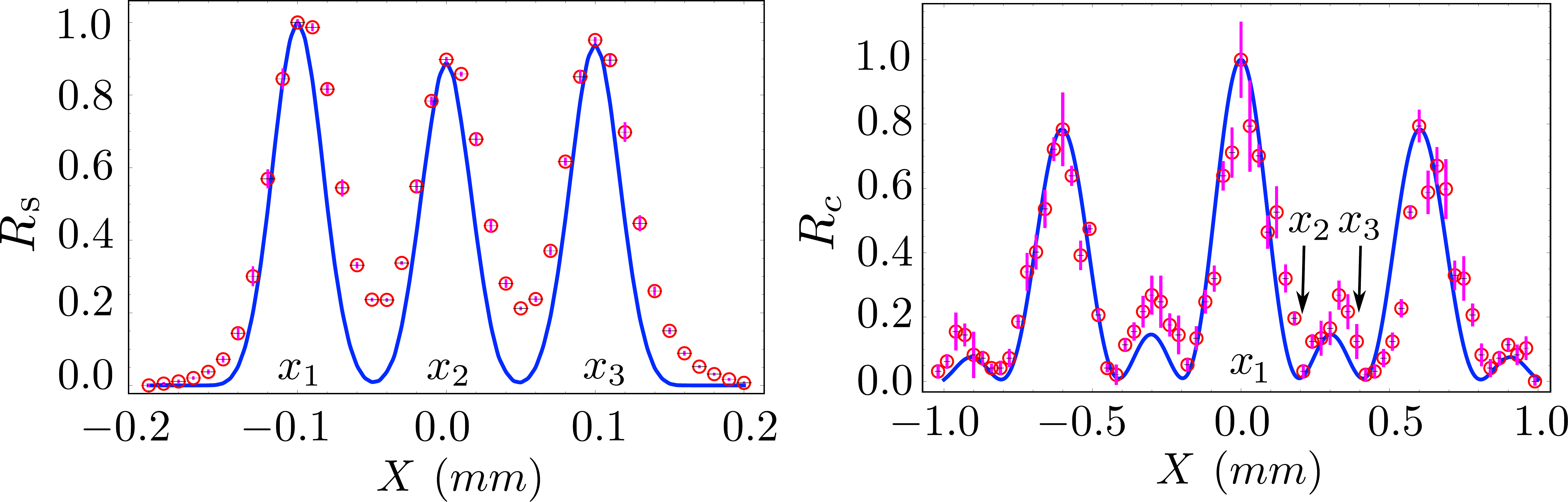}
	\caption{Normalized singles count ($R_s$) vs detector position (X) measured in image plane (left).  Normalized coincidence count ($R_c$) vs detector position (X) measured in focal plane (right). Here, the signal arm detector is fixed at the centre of the singles profile (called $y_1$ in text) while the idler arm is scanned. $x_1, x_2$ and $x_3$ in the plots represent the positions derived from the generalized $\sigma_z$ and  $\sigma_x$ like operators respectively. Theoretical prediction is generated taking the slit widths as $34\mu m$, $28\mu m$ and $32\mu m$ and inter-slit distance as $100\mu m$. }  
\label{fig:combined}
\end{figure*} 
In the current experiment, an SPDC Type-1 BBO source is used to generate spatially correlated spatial-bin qutrit pairs (see Appendix \ref{sectionD} for details). We explore the system (schematic in Fig. 2) at two different detection planes, i.e. focal plane (f) and image plane (2f) of the lenses L2 and L3. When the detectors are placed at the focus of lenses L2 and L3, the lenses transfer the triple-slit interference pattern in the correlation of the signal and idler photons to the detector plane. In addition, there are two cylindrical lenses of focal lengths $50 mm$ and $60 mm$ in the transmitted (let's call it 'signal arm') and reflected (let's call it 'idler arm') arms of beam splitter BS, respectively, which focus the single photons along a line at the detector plane. The individual measured singles spatial profiles of signal and idler photons have a flat top Gaussian profile. \\
We fix one detector at the centre of an individual profile and move the other detector to measure the coincidence. The moving detector is scanned over $2 mm$ in $30\mu m$ step size. At each position, measurement was done thrice, with an accumulation time of $90$ seconds each.\\

Fig. 3 shows the coincidence profile measured in the interference plane. The blue line represents the theoretical prediction while the red circles are the measured values. We include error bars in terms of position (horizontal) and number (vertical) uncertainty. The error in the position is limited by the step size of the actuator which moves the detector. The chosen step sizes of the actuator to measure the profiles are $10\mu m$ and $30\mu m$ for image plane and focal plane respectively. Experimental and theoretically generated Rc  have been appropriately normalised by their respective maxima. The mismatch between experimentally measured counts and the theoretical prediction, especially at the positions of the tertiary maxima in Fig.3, can possibly be denoted to two reasons.\\ Firstly, in the experiment, we use cylindrical lenses before the detectors to focus the profile along a line while the effect of these lenses has not been incorporated in the theory. When a laser beam of radius $r$ is incident on a cylindrical lens, at the focal plane of the cylindrical lens, the circular beam turns into a rectangle with length $4r$ and thickness $2r$ \cite{Powell,Adams}. Thus, the individual peaks of the focused interference profile are rectangles with certain length and thickness depending on the initial radii of the unfocused peaks.  Effectively the detector with $50\times50 \mu m^2$ sensor size can measure more photons from the rectangular area for the small peaks compared to the other.\\ Secondly, the actual slit dimensions are not perfectly captured in the simulations. For instance, the widths are not identical for the three slits. Slit C is the widest at about $34\mu m$, slit B the thinnest at around $28\mu m$ while slit A is in between at around $32\mu m$. We obtain these values by imaging the slits through a microscope and theoretically generate plots using these measured slit widths. Inspite of doing so, there are still some deviations from real scenario like non-smoothness of slit edges as well as finite slit height which contribute to the deviation of experimentally generated profiles from theoretical one.

\subsection{Measurement scheme}

The image plane corresponds to the generalised $\sigma_z$ like operator (see Appendix \ref{sectionE1}) with eigen states comprising of the computational basis states. The three slit peak positions represent the three eigen states with eigen values 0, 1 and -1, respectively as shown in Appendix \ref{sectionE}. In order to calculate the $\mathcal{PCC}$ for generalised $\sigma_z$ like operators applied to both the signal and the idler photons, we need to measure the corresponding joint probabilities. We fix one detector at the three peak positions, one position at a time, of its singles spatial profile (let’s say `signal arm`) and measure the coincidence counts when the other detector is at the peak positions of its singles profile (`idler arm`). Thus, by measuring the peak to peak coincidence counts we construct a 3 x 3 matrix with 9 components. The maximum coincidence counts are the diagonal elements of the matrix. We measure five such sets of matrices. Table I shows a representative correlation matrix. The average $\mathcal{PCC}$ in image plane is calculated from this matrix and found to be 0.904 $\pm$ 0.01. The average $\mathcal{MP}$ in image plane calculated  from the relevant joint probabilities is 0.943 $\pm$ 0.003. \\
The focal plane corresponds to the generalised $\sigma_x$ like operator (see Appendix \ref{sectionE2}) where the eigenstates correspond to three unique positions in the measured cross correlation profile as shown in Appendix \ref{sectionE}. For a pair of observables $\hat{A_1}$ and $\hat{B_1}$ with eigenstates corresponding to the generalised $\hat{\sigma}_{x}$ basis \cite{SGB+06, SHB+12, PRAus} and the quantum state given by $\ket{\psi}=c_{0}\ket{0}\ket{0}+c_{1}\ket{1}\ket{1}+c_{2}\ket{2}\ket{2}$, one can evaluate the quantity $\mathcal{C}_{A_1B_1}$ using Eq. (\ref{pcc6}) for $d=3$.
We need to measure the corresponding joint probabilities like in the case of $\sigma_z$ like operator. To construct such a joint probability matrix, we measure the singles profile for the signal arm and by fitting it with a flat-top Gaussian module function, we find out the centre position naming it $y_1$. Similarly, we find the centre position for the idler singles profile and name it $x_1$. Next, we fix the signal arm at $y_1$ and scan the idler arm to measure the coincidence profile. From this, we extract the positions $x_1$, $x_2$ and $x_3$ corresponding to the eigenstates of the generalised $\sigma_x$ like operator. Next, we fix the idler arm at $x_1$, $x_2$ and $x_3$ respectively and for each fixed position we scan the signal arm to measure the coincidence profiles. From each of the coincidence profiles, we extract the coincidence counts at the positions $y_1$, $y_2$ and $y_3$ which are the derived positions for generalized $\sigma_x$ like operator for signal arm. \\
Thus, we construct a 3x3 correlation matrix. We measure 5 such sets of matrices and Table II shows a representative matrix. We work with operators defined by assigning eigenvalues 0, 1 and -1, respectively to the eigenstates of the $\sigma_x$ like operator and find the average $\mathcal{PCC}$ to be 0.848 $\pm$ 0.027.\\
To evaluate the $\mathcal{MP}$, generalised $\sigma_x$ basis is measured on one side (let's say `signal arm`) and its complex conjugate basis is measured on the other side (`idler arm`). We discuss the case when both observables are the same in Appendix \ref{sectionA1}.
Consider a pure bipartite qutrit state written in Schmidt decomposition as given in Eq.(\ref{mp1}). \\
One can construct the complex conjugate bases of $\{\ket{b_{j}}\}$ as follows
\begin{align}
\label{c1}
\ket{b_{0}}^{*} & = \frac{1}{\sqrt{3}}[\ket{0}+\ket{1}+\ket{2}] \\
\label{c2}
\ket{b_{1}}^{*} & = \frac{1}{\sqrt{3}}[\ket{0}+\omega^{*}\ket{1}+(\omega^{2})^{*}\ket{2}] \\
\label{c3}
\ket{b_{2}}^{*} & = \frac{1}{\sqrt{3}}[\ket{0}+(\omega^{2})^{*}\ket{1}+\omega^{*}\ket{2}] 
\end{align}

where $\omega^{*}$ is the complex conjugate of $\omega$. Using $\omega^{*}=\omega^{2}$ and $(\omega^{2})^{*}=\omega$ we obtain
\begin{align}
\label{c4}
\ket{b_{0}}^{*} & = \ket{b_{0}} \\
\label{c5}
\ket{b_{1}}^{*} & = \ket{b_{2}} \\
\label{c6}
\ket{b_{2}}^{*} & = \ket{b_{1}} 
\end{align}
The above relations imply that one can obtain the probability of detecting the quantum system in the $\ket{b_{0}}^{*}$ or $\ket{b_{1}}^{*}$ or $\ket{b_{2}}^{*}$ state pertaining to the complex conjugate of generalised $\hat{\sigma}_{x}$ basis by using the generalised $\hat{\sigma}_{x}$ basis and obtaining the corresponding probability of detecting the quantum system in $\ket{b_{0}}$ or $\ket{b_{2}}$ or $\ket{b_{1}}$ state, respectively. 

Now, if we consider measuring generalised $\hat{\sigma}_{x}$ operator on one system and its complex conjugate on the other, then by calculating relevant joint probabilities, $P(b_{i},\bar{b}_{j})$ Appendix \ref{sectionA2}, one can obtain \cite{SHB+12} the $\mathcal{MP}$ from Eq. (\ref{mp3}). The quantities $P(b_{1},\bar{b}_{1})$, $P(b_{2},\bar{b}_{2})$ in this case are same as the quantities $P(b_{1},b_{2})$ and $P(b_{2},b_{1})$, respectively, as measured by using generalised $\hat{\sigma}_{x}$ basis on both sides. Thus, summing over the $x_1 - y_1$, $x_2 - y_3$ and $x_3 - y_2$ elements of the correlation matrix as represented in Table II, one can derive $\mathcal{MP}$.  The $\mathcal{MP}$ so derived from correlation matrix comes out to be 0.899 $\pm$ 0.013.\\
Fig.4 represents the detector positions corresponding to generalised $\sigma_z$ and $\sigma_x$ eigenstates, respectively.

\begin{table}
 \begin{tabular}{||c| c| c| c||} 
 \hline
 & $x_1$ & $x_2$ & $x_3$\\ [0.7ex] 
 \hline\hline
 $y_1$ & 0.281 & 0.024 & 0.003 \\ 
 \hline
 $y_2$ & 0.006 & 0.287 & 0.014 \\
 \hline
 $y_3$ & 0.002 & 0.006 & 0.376 \\
 \hline
  \end{tabular}
\caption{3x3 normalized (with respect to the maximum count) correlation matrix corresponding to generalized $\sigma_z$ like operator measured in image plane.}
\end{table}

\begin{table}
 \begin{tabular}{||c| c| c| c||} 
 \hline
 & $x_1$ & $x_2$ & $x_3$\\ [0.7ex] 
 \hline\hline
 $y_1$ & 0.344 & 0.017 & 0.017 \\ 
 \hline
 $y_2$ & 0.008 & 0.017 & 0.260 \\
 \hline
 $y_3$ & 0.017 & 0.302 & 0.017 \\
 \hline
  
\end{tabular}
\caption{3x3 normalized (with respect to the maximum count) correlation matrix corresponding to generalized $\sigma_x$ like operator measured in focal plane}
\end{table}

\section{Results}

The focal plane measurement shown in Fig. 3 has an important significance. In our previous work \cite{Ghosh:18} using the pump beam modulation technique, we demonstrated a high degree of spatial correlation in the image plane. However, the question regarding the essential quantumness of the observed correlations remained unanswered. As discussed in \cite{PhysRevA.57.3123}, when cross correlation measurements (also called coincidence measurements) as a function of detector position in the focal plane for both the signal and idler photons exhibit interference, this implies certification of entanglement.  The current focal plane measurements wherein the normalised coincidence plot exhibits high visibility interference thus answers the previous question by certifying entanglement.\\
Now, we go to the main results of our current work, which is the direct measurement of EM, i.e, $\mathcal{N}$ and $\mathcal{EOF}$. Here one may just briefly remark that for deriving the value of $\mathcal{N}$ from the measured value of $\mathcal{PCC}$, it is empirically advantageous to consider single $\mathcal{PCC}$ because then the associated error range is less than when the sum of $\mathcal{PCC}$s is considered. We calculate $\mathcal{N}$ from the derived $\mathcal{PCC}$ for generalized $\sigma_x$ like operator using Eq. (\ref{pcc6}) and it is 0.848 $\pm$ 0.027. Substituting the derived $\mathcal{MP}$ for the generalized $\sigma_x$ like operator in Eq. (\ref{mp3}) we get $\mathcal{N}$ as 0.848 $\pm$ 0.019. Since for the prepared state, the particular EM given by $\mathcal{N}$ is measured in two different ways by relating it to two different statistical correlators i.e.  $\mathcal{MP}$ and $\mathcal{PCC}$, the value of $\mathcal{N}$, thus measured, being the same within the error range certifies the self-consistency of the procedure employed.\\
Next, for evaluating MI we use Eq. (\ref{MI}) wherein we calculate the $\mathcal{MI}$ where $c_i$ are the normalised coincidence counts when one detector is fixed at $x_i$ (say, $x_1$, then $x_2$ and $x_3$) and the other detector moves from $y_1$ to $y_3$ respectively. $\mathcal{MI}$ is equal to the $\mathcal{EOF}$ in case of computational basis as shown earlier. Our measured $\mathcal{EOF}$ from correlation matrix is 1.233 $\pm$ 0.012. 

\begin{table}
\begin{tabular}{||c| c| c| c||} 
\hline
$\mathcal{N}$ from $\mathcal{PCC}$ & $\mathcal{N}$ from $\mathcal{MP}$ & $\mathcal{EOF}$ from $\mathcal{MI}$ \\  [0.5ex]
\hline
0.848$\pm 0.027$ & 0.849$\pm 0.020$ & 1.233$\pm 0.012$ \\
\hline
\end{tabular}
\caption{Values of different measures of entanglement from correlation matrix.}
\end{table}

\section{Discussion}

The main experimental results are summarised in Table III. We have directly measured two entanglement monotones i.e. $ \mathcal{N}$ and $\mathcal{EOF}$ by using their analytically derived relations with statistical correlation measures i.e. $\mathcal{PCC}, \mathcal{MP}$ and $\mathcal{MI}$. 

\subsection{Purity of our experimentally generated state}

The purity of the experimentally generated state is a critical assumption towards using our analytical relations. We thus go on to demonstrate through multiple arguments below, the reasonableness of this assumption.

We know that the concept of Coherence is intimately connected with mixedness of the state \cite{PhysRevA.92.022316} and a good measure of coherence is the Visibility of the interference \cite{PhysRevLett.85.2845}. Higher degree of coherence implies higher Purity of the state (or lowed mixedness). In our experiment, the interference in the coincidence plane has a Visibility of $\sim 91.7 \%$ which indicates a high Purity of the state. The translation stage that we had access to for measuring the interference pattern has a precision of 10 $\mu m$. With higher precision stages, we would have likely achieved even higher visibility.

Now, in order to further substantiate the above argument for high Purity of the prepared state from different considerations, let us make an ansatz that we have a small amount of white noise added to our pure state which turns it into an isotropic mixed state. The density matrix of such a state is, (for detailed derivation please see \ref{sectionB} and Appendix \ref{sectionC}.)

\begin{equation}
\label{d1}
\rho=\alpha\ket{\psi_{AB}}\bra{\psi_{AB}}+\frac{1-\alpha}{d^2}\hat{\mathbb{I}}
\end{equation} 
where $\ket{\psi_{AB}}$ is the pure two-qudit state given by Eq. (\ref{mp1}). If our original pure state- assumption is to be verified, then it is required that the mixedness parameter $\alpha=1$. How do we check that? We do so by relating the experimentally measured $\mathcal{MP}$ with $\alpha$ and $\mathcal{N}$ respectively for the state given by Eq. (\ref{d1}).
Calculating relevant joint probabilities the $\mathcal{MP}$ for the state for $\sigma_x$ basis is

\begin{equation}
\label{d2}
\mathcal{MP}=\sum^{d-1}_{i=0}P(A,B)=\alpha+\frac{1-\alpha}{d}
\end{equation}

For isotropic mixed state the $\mathcal{MP}$ is related to $\mathcal{N}$ as follows,
\begin{equation}
\label{d3}
\mathcal{MP}=\sum^{m}_{i=1}P(A,B)=\frac{2m}{d+1} (\mathcal{N}+1)
\end{equation}
where $m$ is the number of MUBs.\\
The $\mathcal{MP}$ derived from correlation matrix in the focal plane measurement comes out to be 0.899 $\pm$ 0.013. Putting the derived $\mathcal{MP}$ in Eq. (\ref{d3}) for $m=1$ and $d=3$, $\mathcal{N}$ comes out to be 0.798 $\pm$ 0.026. From Eq. (\ref{d2}) we get $\alpha$ as 0.848 $\pm$ 0.019 showing that the probability of our state being a pure maximally entangled state is $\sim 85 \%$. Thus, the high values of Purity of the prepared state inferred from the preceding two independent lines of argument suggest that if we evaluate the value of $\mathcal{N}$ from the experimentally determined $\mathcal{PCC}$ or $\mathcal{MP}$ assuming the prepared sate to be pure, it should be quite close to the value of $\mathcal{N}$ determined assuming the state to be isotropic mixed. That this is so is corroborated as follows. \\
By invoking Eq. (\ref{pcc6}) we get $\mathcal{N}$ from calculated $\mathcal{PCC}$ from correlation matrix measured in the focal plane to be 0.848 $\pm$ 0.019 and the $\mathcal{N}$ from $\mathcal{MP}$ by using Eq. (\ref{d3}) equal to 0.798 $\pm$ 0.026. Therefore, the values of $\mathcal{N}$ that we derive from both pure state assumption as well as with the presence of small amount of white noise assumption are almost the same within the error bars. \\
We also find out the $\mathcal{N}$ from the derived $\mathcal{MP}$ using two MUBs, generalised $\sigma_z$  basis is measured on both sides and generalised $\sigma_x$ basis is measured on one side and its complex conjugate on the other side. Using Eq. (\ref{d3}) putting $m=2$, from the sum of the $\mathcal{MP}$s the calculated $\mathcal{N}$ is 0.842 $\pm$ 0.029 and using Eq. (\ref{pcc8}), considering the $\mathcal{MP}$ derived for computational basis is $1$. So, $\mathcal{N}$ is 0.752 $\pm$ 0.029. Thus they are close within the error bound proving that even when we use two MUBs, our hypothesis that the state is pure seems reasonable. \\
Finally, a further validation of the internal consistency of the preceding line of analysis of our experimental results is provided by the sameness (within the respective error bars) of the values of $\mathcal{N}$ derived from $\mathcal{PCC}$ and $\mathcal{MP}$ respectively, taking the prepared state to be pure.\\

\subsection{Quantitative comparison of  entanglement monotones}

Here we consider the following operationally significant question: what is the percentage deviation of the entanglement of the prepared state from the maximally entangled state as quantified by two different entanglement measures, i.e.,  $\mathcal{N}$ and $\mathcal{EOF}$? Noting that the maximum values of $\mathcal{N}$ and $\mathcal{EOF}$ are 1 and $\log_2 3=1.585$, respectively, it is found that while our measured $\mathcal{N}$  value quantifies a $\sim 15.2 \%$ deviation from the maximally entangled state, the deviation captured by our measured $\mathcal{EOF}$ is around $\sim 22.2 \%$. Thus, the appreciable difference in these percentage deviations demonstrates non-equivalence between different measures of entanglement in higher dimensions. In an earlier study, this feature of non-equivalence has been analysed by us for two-qubit pure state \cite{2019arXiv190709268S}. Ours is the first experimental demonstration of  non-equivalence (in the sense defined above), between different measures of entanglement in higher dimensional bipartite pure states for a prepared two-qutrit state.
We now provide a generalised illustration of this feature for an arbitrary two-qutrit pure state by considering different choices of Schmidt coefficients.

Consider a two qutrit pure state with Schmidt coefficients $c_{0}$ , $c_{1}$ \& $c_{2}$ 
\begin{align}
\label{qutrit}
    \ket{\Phi} = c_{0}\ket{0}\ket{0} + c_{1}\ket{1}\ket{1} + c_{2}\ket{2}\ket{2} 
\end{align}
where $0 < c_{0}$,$c_{1},c_{2} < 1$ and $c_{0}^2$ + $c_{1}^2$  + $c_{2}^2$ = 1

Two parameters are defined to measure the percentage deviations of measures from  the values corresponding to Maximally Entangled State \cite{2019arXiv190709268S}

\begin{align}
Q_{E} = ((\log_{2}(3) - E)/\log_{2}(3))\times 100 
\end{align}
\begin{align}
 Q_{N} = (1 - N)\times 100    
\end{align}

To see to what extent these two parameters differ with each other, the following quantity is an appropriate measure \cite{2019arXiv190709268S}
\begin{align}
 \Delta Q_{NE} = |Q_{E}- Q_{N}|
 \end{align}

\begin{itemize}
    \item Different values of $\Delta Q$ for different values of the Schmidt coefficients have been tabulated in Table I. 
    \begin{table}[h]
    \begin{center}
    \caption{Differences in the \% deviations from the value corresponding to the maximally entangled state}
    \label{tab:table1}
    \begin{tabular}{l|c|c|c|c|c|r} 
      ${c_{0}}$ & ${c_{1}}$ & \textbf{E} & \textbf{N} & ${Q_{E}}$ & ${Q_{N}}$  & {$\Delta Q_{NE}$}\\
      
      \hline
      
      0.1 & 0.1 & 0.1614 & 0.2080 & 89.8142 & 79.2010 & 10.6132 \\
      0.3 & 0.8 & 1.2347 & 0.8116 & 22.0964 & 18.8423 & 3.2540 \\
      0.5774 & 0.5774 & 1.5850 & 1 & 0 & 0 & 0 \\
      0.6 & 0.6 & 1.5755  & 0.9950  & 0.6001 & 0.5020 & 0.0982\\
      0.9 & 0.3 & 0.8911 & 0.6495 & 43.7784 & 35.0527 & 8.7257\\

    \end{tabular}
     \end{center}
    \end{table}
    
Observations: 
     \item Given a non-maximally entangled two qutrit pure state, one cannot comment as in the case of two qubit case \cite{2019arXiv190709268S} that one entanglement measure is always greater than other entanglement measure for any value of state parameter. 
     
     \item $\Delta Q_{NE}$ takes a maximum value of 12.148\% when $c_{0} = 0.1712$ \&  $c_{1} = 0.1712$ 
    \item Thus, both as absolute and relative entanglement measures, Negativity and Entanglement of Formation do not give equivalent results. 

\end{itemize}

\subsection{Non-monotonicity between $\mathcal{N}$ and Entanglement of Formation ($\mathcal{E}$) for two qutrit pure states}

For two-qubit pure states, it is known the different entanglement measures are monotonically related \cite{2019arXiv190709268S}. In contrast, in higher dimensional systems, studies have suggested that for certain classes of states, the different measures of entanglement are not monotonically related to each other \cite{MG04, CS08, Ero15}. In other words, whether one state is more entangled than the other will depend on the choice of the entanglement measure! This opens up interesting questions regarding the optimal choice of entanglement measure in different higher dimensional quantum information protocols and calls for deeper understanding of the meanings of the respective entanglement measures for their meaningful comparison. Here we illustrate this non-monotonicity for an arbitrary two-qutrit pure state as defined in Eq. (\ref{qutrit}) by considering different choices of Schmidt coefficients.\\

Rate of change of $\mathcal{E}$ w.r.t $c_{0}$
\begin{equation}
\dfrac{\mathrm{d}\mathcal{E}}{\mathrm{d}c_{0}}= (2/ln(2))c_{0}\log_{2}(1-(c_{0}^2+c_{1}^2))/c_{0}^2) \label{m1}
\end{equation}
Similarly, rate of change of $\mathcal{E}$ w.r.t $c_{1}$
\begin{equation}
   \dfrac{\mathrm{d}\mathcal{E}}{\mathrm{d}c_{1}} = (2/ln(2))c_{1}\log_{2}(1-(c_{0}^2+c_{1}^2))/c_{1}^2) \label{m2}
\end{equation}
 Rate of change of $\mathcal{N}$ w.r.t $c_{0}$
\begin{equation}
   \dfrac{\mathrm{d}\mathcal{N}}{\mathrm{d}c_{0}} = c_{1} + (1-c_{0}c_{1}-c_{1}^2-2c_{0}^2)/\sqrt{1-(c_{0}^2+c_{1}^2)} \label{m3}
\end{equation}
Similarly, rate of change of $\mathcal{N}$ w.r.t $c_{1}$
\begin{equation}
   \dfrac{\mathrm{d}\mathcal{N}}{\mathrm{d}c_{1}} = c_{0} + (1-c_{0}c_{1}-c_{0}^2-2c_{1}^2)/\sqrt{1-(c_{0}^2+c_{1}^2)} \label{m4}
\end{equation}
Observations:
\begin{itemize}
    \item From Eqs. (\ref{m1}),(\ref{m2}),(\ref{m3}),(\ref{m4}) it can be seen that for a given $c_{0}$ ($c_{1}$), $\mathcal{E}$ and $\mathcal{N}$ grow with $c_{0}$ ($c_{1}$), reach a certain value and then start decreasing w.r.t $c_{0}$ ( $c_{1}$). All the above four Eqs.(\ref{m1}),(\ref{m2}),(\ref{m3}),(\ref{m4}) vanish when $c_{0}$ = $c_{1}$ = $1/\sqrt{3}$
    
    \item As in the case of two qubit pure state \cite{2019arXiv190709268S}, one cannot say that $\dfrac{\mathrm{d}\mathcal{E}}{\mathrm{d}\mathcal{N}}$ is always greater than zero except when $c_{0}$ \& $c_{1}$ = $1/\sqrt{3}$. This explains the presence of non-monotonic nature between these two parameters. For example, consider a pair of two qutrit pure states with Schmidt coefficients $c_{0}$ = 0.4 , $c_{1}$ = 0.9 \& $c_{0}$ = 0.5 , $c_{1}$ = 0.1. Former state has, 
    $$ E_{1} = 0.8879\  \&\  N_{1} = 0.5661 $$
    whereas the latter state has 
    $$ E_{2} = 0.8210\  \&\  N_{2} = 0.5852 $$
    Here $E_{1} > E_{2}$ but $N_{1} < N_{2}$, showing $\mathcal{E}$ and $\mathcal{N}$ are not monotonic w.r.t each other.

\end{itemize}


\section*{Conclusion}

In conclusion, an experimental procedure is formulated to measure different entanglement monotones with limited number of measurements. A proof-of-concept experimental demonstration is provided in an architecture consisting of pump beam modulated spatially correlated bipartite qutrits. This therefore sets up, for the first time, a platform enabling the direct determination of and comparison of the EMs for the higher dimensional entangled states, based on only one set of joint local complementary measurements. 
In particular, using this platform, a follow-up work could be to demonstrate the non-monotonicity between the different EMs for the two-qutrit pure states by suitably preparing different entangled states, whose theoretical basis has been discussed in the present paper. Since, for the two-qubit pure entangled states, the EMs are known to be monotonic with respect to each other, this makes the study of the deviation from this feature especially interesting for the higher dimensional entangled states. It is worth emphasizing here that although for the mixed states, such non-monotonicity between the EMs is well known both for the two-qubit as well as for the two-qudit entangled states, for the pure states, this feature has surprisingly remained uninvestigated for the higher dimensional case.  
Further, since in our work the EMs have been linearly related with the relevant statistical correlators, such non-monotonicity between the EMs would imply non-monotonicity between the statistical correlators, too. This, in turn, would mean that, given two pure entangled two-qudit states, whether any one of them can be regarded as more entangled/correlated than the other would depend upon which EM/statistical correlator is used. Thus, the import of this finding can have ramifications concerning the deeper meanings of both EMs and statistical correlators in the context of quantifying correlations in the higher dimensional bipartite pure entangled states. \\ 

\section*{Acknowledgments}
The authors thank D.Home, C. Jebarathinam, S. Kanjilal, P. Kolenderski, S. Sadana and A. Sinha for useful discussions on theoretical concepts. Thanks are due to S. Bhar for initial discussions.

\section*{Data Availability}
The data that support the plots in this paper and other details of this study are available from the corresponding author on reasonable request.

\clearpage

\onecolumngrid

\appendix

\input{appendix.tex}
\end{document}

%% file: appendix.tex
\section{Relating Mutual Predictability with Negativity}
\label{sectionA}
Consider a pure bipartite state written in Schmidt decomposition,
\begin{equation}
\label{s1}   
\psi_{AB}=\sum_{i=0}^{d-1} \lambda_i \ket{i_A} \ket{i_B}
\end{equation} with density matrix,
\begin{equation}
\label{density}
\rho_{AB}=\sum\limits_{i,j=0}^{d-1} \lambda_i \lambda_j \ket{i_A,i_B}\bra{j_A,j_B}
\end{equation} where $\{\ket{i_A,i_B}\}$ is Schmidt basis for the state.
Mutual Predictability ($\mathcal{MP}$) w.r.t the measurement basis, $\{\ket{\psi_m,\phi_n}\})$ is, where $\{\ket{\psi_m}\}$ and $\{\ket{\phi_n}\}$ be the measurement bases for sub-systems A and B respectively.
\begin{equation}
\label{s2}
\mathcal{MP}=\sum_{m=0}^{d-1}\, \sum_{i,j=0}^{d-1} \lambda_i \lambda_j  \braket{\psi_m ,\phi_m \vert a_i, b_i} \braket{a_j,b_j \vert \psi_m,\phi_m}
\end{equation}
To quantify the entanglement via $\mathcal{MP}$ we calculate it w.r.t  $k^{th}$ MUB. Let's consider the set of MUB, $\ket{\psi_m^k,\phi_n^k}$ w.r.t the arbitrary measurement basis, $\ket{\psi_a,\phi_b}$.
\begin{align*}
     \mathcal{MP}^k  =  \sum_{m=0}^{d-1} \sum_{i,j=0}^{d-1} \lambda_i \lambda_j \braket{\psi_m^k, \phi_m^k | i_A,i_B}\braket{j_A,j_B | \psi_m^k, \phi_m^k } \\
      = \sum_{a,b,c,d=0}^{d-1}\sum_{m=0}^{d-1} \sum_{i,j=0}^{d-1} \lambda_i \lambda_j 
     \braket{\psi_m^k, \phi_m^k | \psi_a, \phi_b}\braket{\psi_a, \phi_b | i_A,i_B}\\
     \times \braket{j_A,j_B | \psi_c, \phi_d}\braket{\psi_c, \phi_d | \psi_m^k, \phi_m^k } 
\end{align*}

\begin{align*}
 \sum_{i,j}^{}\lambda_i\lambda_j\braket{\psi_a,\phi_b \vert i_A,i_B}\braket{j_A,j_B \vert \psi_c,\phi_d}\leq \lambda_a \lambda_c \delta_{ab}\delta_{cd}   
\end{align*}
Consequently, 
\begin{align*}
\mathcal{MP}^k &\leq \sum_{m=0}^{d-1}\sum_{a,c}^{}\lambda_a \lambda_c\braket{\psi_m^k,\phi_m^k\vert \psi_a\phi_b}\braket{\psi_c\phi_d\vert\psi_m^k,\phi_m^k}\\
& \leq \sum_{m=0}^{d-1}\sum_{a,c}^{}\lambda_a \lambda_c\vert \braket{\psi_m^k,\phi_m^k\vert \psi_a\phi_b}\vert^2 \\
 & \leq \sum_{a,c}^{}\frac{\lambda_a \lambda_c}{d}
\end{align*}
where $\vert \braket{\psi_m^k,\phi_m^k\vert \psi_a,\phi_b}\vert=\frac{1}{\sqrt{d}}$.\\
So, the maximum value of sum of $\mathcal{MP}$ over the optimal set of MUB is,
\begin{equation}
\label{s4}
\begin{split}
\mathcal{MP}_k^{max}=\sum_{a,c}^{}\frac{\lambda_a \lambda_c}{d}\\
=\frac{1+2\mathcal{N}}{d}
\end{split}
\end{equation}

\subsection{Mutual Predictability when generalized $\sigma_{x}$ observable is measured on both sides}
\label{sectionA1}
Consider a pure bipartite qutrit state written in Schmidt decomposition as given in Eq.(\ref{s1}). \\
Let the basis $\{\ket{b_{j}}\}$ be the generalized $\sigma_x$ basis \cite{SGB+06, SHB+12, PRAus}
\begin{align}
\label{sigmax1}
\ket{b_{0}} & = \frac{1}{\sqrt{3}}[\ket{0}+\ket{1}+\ket{2}] \\
\label{sigmax2}
\ket{b_{1}} & = \frac{1}{\sqrt{3}}[\ket{0}+\omega\ket{1}+\omega^{2}\ket{2}] \\
\label{sigmax3}
\ket{b_{2}} & = \frac{1}{\sqrt{3}}[\ket{0}+\omega^{2}\ket{1}+\omega\ket{2}] 
\end{align}
where $\omega=e^{2i\pi/3}$ with $i=\sqrt{-1}$.
\begin{equation}
\label{sigmax4}
P(b_{i},b_{j})=|\bra{\psi}\ket{b_{i}}\bra{b_{i}}\otimes\ket{b_{j}}\bra{b_{j}}\ket{\psi}|^{2}
\end{equation}
Using Eq.(\ref{sigmax4}) we obtain the following joint probabilities

\begin{align}
\label{ss1}
P(b_{0},b_{0}) & = P(b_{1},b_{2})=P(b_{2},b_{1})\\&=\frac{1}{9}(1+2c_{0}c_{1}+2c_{2}c_{1}+2c_{0}c_{2})\\
\label{ss2}
P(b_{1},b_{1}) & = P(b_{2},b_{2})=P(b_{0},b_{2})\\&=P(b_{2},b_{0})=P(b_{0},b_{1})=P(b_{1},b_{0})\\&=\frac{1}{9}(1-c_{0}c_{1}-c_{2}c_{1}-c_{0}c_{2})
\end{align}
It can be checked from Eqs.(\ref{ss1}) and (\ref{ss2}) that $\sum_{i,j}P(b_{i},b_{j})=1$.\\
Using the relevant joint probabilities from Eqs.(\ref{ss1}) and (\ref{ss2}) one can then obtain Mutual Predictability \cite{SHB+12} as 
\begin{align}
\label{MPeqn1}
\mathcal{MP} & = \sum_{i}P(b_{i},b_{i})\\
\label{MPeqn2}
& = \frac{1}{3}
\end{align}

\subsection{Mutual Predictability when generalized $\sigma_{x}$ observable is measured on one side and its complex conjugate is measured on the other side}
\label{sectionA2}

Consider a pure bipartite qutrit state written in Schmidt decomposition as given in Eq.(\ref{s1}). \\
One can construct the complex conjugate bases of $\{\ket{b_{j}}\}$ as follows
\begin{align}
\label{conj1}
\ket{b_{0}}^{*} & = \frac{1}{\sqrt{3}}[\ket{0}+\ket{1}+\ket{2}] \\
\label{conj2}
\ket{b_{1}}^{*} & = \frac{1}{\sqrt{3}}[\ket{0}+\omega^{*}\ket{1}+(\omega^{2})^{*}\ket{2}] \\
\label{conj3}
\ket{b_{2}}^{*} & = \frac{1}{\sqrt{3}}[\ket{0}+(\omega^{2})^{*}\ket{1}+\omega^{*}\ket{2}] 
\end{align}

where $\omega^{*}$ is the complex conjugate of $\omega$. Using $\omega^{*}=\omega^{2}$ and $(\omega^{2})^{*}=\omega$ we obtain
\begin{align}
\label{conj4}
\ket{b_{0}}^{*} & = \ket{b_{0}} \\
\label{conj5}
\ket{b_{1}}^{*} & = \ket{b_{2}} \\
\label{conj6}
\ket{b_{2}}^{*} & = \ket{b_{1}} 
\end{align}

Eqs.(\ref{conj4})-(\ref{conj6}) imply that one can obtain the probability of detecting the quantum state in the $\ket{b_{0}}^{*}$ or $\ket{b_{1}}^{*}$ or $\ket{b_{2}}^{*}$ state pertaining to the complex conjugate of generalized $\sigma_x$ basis by using the generalized $\sigma_x$ basis and obtain the corresponding probability of detecting the quantum state in $\ket{b_{0}}$ or $\ket{b_{2}}$ or $\ket{b_{1}}$ state respectively.\\
Now, if we consider measuring generalized $\sigma_x$ on one side and its complex conjugate on the other side, then we can obtain the joint probabilities $P(b_{i},\bar{b}_{j})$ as
\begin{equation}
\label{both1}
P(b_{i},\bar{b}_{j})=|\bra{\psi}\ket{b_{i}}\bra{b_{i}}\otimes\ket{b_{j}^{*}}\bra{b_{j}^{*}}\ket{\psi}|^{2}
\end{equation}
where $\bar{b}_{j}$ denotes the eigenvalue corresponding to the complex conjugate of the $\ket{b_{j}}$, whence we obtain
\begin{align}
\label{both2}
P(b_{0},\bar{b}_{0}) & = P(b_{1},\bar{b}_{1})=P(b_{2},\bar{b}_{2})=\frac{1}{9}(1+2c_{0}c_{1}+2c_{2}c_{1}+2c_{0}c_{2})
\end{align}
As discussed in the preceding paragraph, the quantities $P(b_{1},\bar{b}_{1})$, $P(b_{2},\bar{b}_{2})$ in this case are same as the quantities $P(b_{1},b_{2})$ and $P(b_{2},b_{1})$ given by Eq. (\ref{ss1}) respectively, as measured by using generalized $\sigma_x$ basis on both sides.\\
Now, using Eq. (\ref{both2}), one can obtain \cite{SHB+12} the Mutual Predictability as 
\begin{align}
\label{}
\mathcal{C} & = \sum_{i}P(b_{i},\bar{b}_{i})\\
\label{MPeqn3}
& = \frac{1}{3}(1+2c_{0}c_{1}+2c_{2}c_{1}+2c_{0}c_{2})\\
\label{MPeqn4}
& = \frac{1}{3}(1+2\mathcal{N})
\end{align}
where $\mathcal{N}$ is the negativity of the bipartite qutrit state \cite{ETS15}.

\section{ Details on the photon source and pump beam modulation technique}
\label{sectionD}

A Type-1, non-linear crystal (BBO) with a dimension of 5 mm x 5 mm x 10 mm generates parametrically down-converted degenerate photons at 810 nm wavelength with collinear phase matching condition. A diode laser at 405 nm with 100 mw power pumps the crystal. The transverse spatial profile of the pump beam at the crystal is prepared by transferring the laser beam
through a three-slit aperture with 30 $\mu m$ slit width and 100  $\mu m$ inter-slit distance and imaging it at the crystal. A plano-convex lens (L1) of focal length 150 mm is placed such that it forms the image of the three-slit at the crystal. A 50-50 beam-splitter (BS) placed after the crystal splits the two down-converted photons in the transmitted and reflected ports of the BS. A band-pass filter centred at 810 nm with a FWHM of 10 nm passes the down-converted photons and a long-pass filter with cut-off wavelength 715 nm blocks the residual pump. In each arm of the BS, a plano-convex lens (L2 and L3 respectively) of focal length 75 mm is placed at 2f distance from the crystal to transfer the signal and idler photon profile to the detectors.

\section{Experimentally defining generalized $\sigma_x$ and generalized $\sigma_z$ bases}
\label{sectionE}

\subsection{Experimental realization of $\sigma_z$}
\label{sectionE1}

The state of the down-converted photon after passing through a three-slit can be written as, 
\begin{equation} 
\ket{\psi} = \dfrac{1}{\sqrt{3}} ( c_0 \ket{0} + c_1 \ket{1} + c_2 \ket{2}) \label{exp1}
\end{equation}
The probability to detect a photon prepared in the state $\ket{\psi}$ in the position  corresponding to the $n$-th
slit image is proportional to $|{c_n}|^2$ 
and the measurement operators can be defined as  
\begin{equation}
M_{nf}(n)= \mu_{nf} \ket{n}\bra{n} \label{exp2}
\end{equation}
where $\mu_{nf} $ is the normalization factor\cite{PhysRevA.86.012321}.
The $\sigma_z$ matrix in 3-dimension is 
\begin{equation}
\sigma_z = \begin{bmatrix} 1 & 0 & 0\\ 0 & 0 & 0 \\ 0 & 0 & -1  	\end{bmatrix}
\end{equation}
which can be written as
\begin{equation}
\sigma_z = \ket{0}\bra{0} - \ket{2}\bra{2} \label{exp3} 
\end{equation}
So the positions corresponding to the eigen bases are the center of the first and third slit image profile.

\subsection{Experimental realization of $ \sigma_x$}
\label{sectionE2}

A detection in the position $x$ in the focal plane corresponds to the projector onto $\ket{k_x}$. 
\begin{equation}
k_x =   xk /f \label{exp4}
\end{equation} 
where $k_x$ is the transverse wave vector, $k$ is the wave number and $f$ is the focal length of the lens. The detection probability is proportional to \cite{PhysRevA.86.012321}
\begin{equation}
 |\sqrt{\dfrac{a}{2\pi}} {\rm sinc}(k_x a/2) \ket{\phi(k_x d)}\bra{\psi}|^2  \label{exp5}
\end{equation} 
where $ \ket{\phi(\theta)} = \ket{0} + \exp{i \theta}\ket{1}  + \exp{2i \theta}\ket{2}$. Hence, the measurement operators in the far field can be defined as $M_{ff}(\theta) \, = \mu_{ff}\, \ket{\phi(\theta)}\bra{\phi(\theta)}$ and the phase parameter is 

\begin{equation}
\theta = 2 \pi x d /\lambda f \label{exp6}
\end{equation}

the operator $ \sigma_x$ can be written as

\begin{equation}
O_1 = \ket{b_0}\bra{b_0} -\ket{b_2}\bra{b_2} \label{exp7}
\end{equation}
where $\ket{b_0}$,$\ket{b_1}$ and $\ket{b_2}$ are 
\begin{equation}
\ket{b_0} =  \dfrac{1}{\sqrt{3}} (\ket{0} + \ket{1} + \ket{2}) \label{exp8}
\end{equation}
\begin{equation}
\ket{b_1} = \dfrac{1}{\sqrt{3}} (\ket{0} + e^{2i\pi/3}\ket{1} + e^{-2i\pi/3}\ket{2}) \label{exp9}
\end{equation}
\begin{equation}
\ket{b_2}= \dfrac{1}{\sqrt{3}} (\ket{0} + e^{-2i\pi/3}\ket{1} + e^{2i\pi/3}\ket{2}) \label{exp10}
\end{equation}
The corresponding angles of the eigenvectors of $\sigma_x$ are ${0, \dfrac{2 \pi}{3}  and \dfrac{4 \pi}{3}}$ respectively. The corresponding detector positions are $0$, $202.5$ and $405$ $\mu m$ which are $x_1$,  $x_2$ and  $x_3$ as mentioned in main paper; where, $\lambda = 0.810 \mu m, f = 7.5 cm, d = 100 \mu m$.

\section{Relating Purity with Mutual Predictability for an isotropic mixed state}
\label{sectionB}

A pure maximally entangled bipartite state can be written as
\begin{equation}
\ket{\psi_d}=\frac{1}{\sqrt{d}}\sum^{d-1}_{i=0}\ket{i,i} \label{sup1}
\end{equation}
Some mixedness can be introduced due to white noise and such a mixed state takes the form of an isotropic state with density matrix
\begin{equation}
\rho=\alpha\ket{\psi_d}\bra{\psi_d}+\frac{1-\alpha}{d^2} \mathbb{I} ,   \label{sup2}
\end{equation} 
where $0\le\alpha\le1$, the state is pure if $\alpha=1$.
The joint probability of outcomes in the computational basis for an isotropic state is 

\begin{equation}
\begin{aligned}
P(i_A,i_B)=\bra{i_A,i_B}\rho\ket{i_A,i_B}   \\ 
=\frac{\alpha}{d}\sum^{d-1}_{m,n=0}\bra{i_A,i_B}\ket{m,m}\bra{n,n}\ket{i_A,i_B}+\frac{1-\alpha}{d^2} \\
=\delta_{i_A,i_B}\frac{\alpha}{d}+\frac{1-\alpha}{d^2}  \label{sup3} 
\end{aligned}
\end{equation} 
where $i_A$ and $i_B$ are the outcomes of measurement on the respective subsystems. If the measurement basis for subsystem A and B is complex conjugate of each other then the joint probability remains the same under any unitary transformed measurement basis chosen, as isotropic state is $\mathbb{U}\otimes\mathbb{U^{*}}$ invariant. In
the manuscript the measurement basis chosen is the $\sigma_x$ basis for A and its complex conjugate for B. Therefore, the expression for the joint probability holds for such measurements. $\mathcal{MP}$ for the isotropic state for $\sigma_x$ basis is
\begin{equation}
C_{A_i,B_i}=\sum^{d-1}_{i=1}P(i,i)=\alpha+\frac{1-\alpha}{d}   \label{sup4}
\end{equation}

\section{Relating Mutual Predictability with Negativity for an isotropic mixed state}
\label{sectionC}

The density matrix of an isotropic mixed state can be written alternatively in terms of fidelity $F$,
\begin{equation}
\rho_F=\frac{d^2}{d^2-1}\left[\frac{1-F}{d^2}\mathbb{I}+\left(F-\frac{1}{d^2}\right)\ket{\psi_d}\bra{\psi_d}\right] \label{sup5}
\end{equation}
Then we can write Eq.(\ref{sup4}) as following,
\begin{equation}
C_{A_i,B_i}=\sum^{d-1}_{i=1}\frac{1}{d^2}\left(1+\frac{d^2F-1}{d+1}\right)=\frac{1}{d}\left(1+\frac{d^2F-1}{d+1}\right)  \label{sup6}
\end{equation}
Let $m\le d$ be the number of existing MUBs. Then the sum of $\mathcal{MP}$ in all MUBs is
\begin{equation}
\sum^{m}_{i=1}C_{A_i,B_i}=\frac{m}{d}\left(1+\frac{d^2F-1}{d+1}\right)=\frac{m}{d+1}\left(1+dF\right)    \label{sup7}
\end{equation}
As the $\mathcal{N}$ for an isotropic state is 
\begin{equation}
\mathcal{N}(\rho_F)=max\left(\frac{dF-1}{2},0\right)  \label{sup8}
\end{equation}
the sum of $\mathcal{MP}$s related to the $\mathcal{N}$ for $ F> \frac{1}{d}$ as
\begin{equation}
\sum^{m}_{i=1}C_{A_i,B_i}=\frac{2m}{d+1}(\mathcal{N}+1)   \label{sup9}
\end{equation}
For the special case, $m=d+1$ Eq.(\ref{sup9}) simplifies to
\begin{equation}
\sum^{m}_{i=1}C_{A_i,B_i}=2(\mathcal{N}+1)   \label{sup10}
\end{equation}